\documentclass[aps,prl,floatfix,showpacs,twocolumn]{revtex4}
\usepackage{amsmath,amssymb}
\usepackage{graphicx}
\usepackage{epsfig}
\usepackage{psfrag}

\newcommand{\bm}{\boldsymbol}

\begin{document}

\hsize\textwidth\columnwidth\hsize\csname@twocolumnfalse\endcsname

\title{Phonon-induced many-body renormalization of graphene electronic properties}

\author{Wang-Kong Tse}
\author{S. Das Sarma}
\affiliation{Condensed Matter Theory Center, Department of Physics,
University of Maryland, College Park, Maryland 20742}

\begin{abstract}
We develop a theory for the electron-phonon interaction effects on the electronic properties of graphene. We analytically calculate the electron self-energy, spectral function and band velocity renormalization due to phonon-mediated electron-electron interaction. We find that phonon-mediated electron-electron coupling has a large effect on the graphene band structure renormalization, and our analytic theory successfully captures the essential features of the observed graphene electron spectra in the ARPES experiments, predicting a kink at $\sim 200\mathrm{meV}$ below the Fermi level and a reduction of the band velocity by $\sim 10-20\%$ at the experimental doping level. 
\end{abstract}

\maketitle
\newpage
Electronic properties of graphene are of fundamental importance because of its linear two-dimensional Dirac-like energy dispersion, attracting considerable current attention \cite{SSC}. Although there has been a great deal of theoretical work on the effects of disorder and electron-electron (e-e) interaction on graphene electronic properties, the effect of electron-phonon (e-ph) interaction in graphene has not been studied extensively. In this Letter, we present a leading order many-body theoretic analysis of e-ph interaction-induced renormalization of graphene electronic properties, comparing our theoretical results critically with existing experimental results \cite{ARPES}. The motivation of our theoretical work comes from the beautiful recent experimental ARPES (angle-resolved photoemission) study \cite{ARPES} of graphene electronic spectrum, and our theory explains the main features of the observed ARPES spectra as arising from the e-ph interaction.

We focus our investigation on the effect of e-ph interactions on band structure renormalization by calculating the electron self-energy due to the phonon-mediated e-e interaction. Graphene has a two-dimensional honeycomb real-space lattice structure comprising two interpenetrating triangular sublattices A and B, which translates to a reciprocal-space honeycomb structure with the hexagonal Brillouin zone cornering at the high-symmetry K points. In the vicinity of these K points (so-called Dirac points), the low-energy excitations have a linear energy spectrum described by the low-energy bare Hamiltonian $H_0 = v\bm{\sigma}\cdot\bm{k}$, where $v \approx 10^6\mathrm{ms}^{-1}$ is the Fermi velocity of the Dirac 
fermions, $\bm{\sigma}$ is the set of Pauli matrices 
representing the two (A and B) sublattice ``pseudo-spin'' degrees of freedom 
 (we let $\hbar = 1$ throughout unless it is written out explicitly). This Hamiltonian 
describes a cone-like linear energy spectrum with $\epsilon_{k\lambda} = \lambda \epsilon_k$, where $\epsilon_k = vk$, 
$\lambda$ is the chirality label standing for the conduction band ($\lambda = 1$) or the valence band ($\lambda = -1$). 
The e-ph interaction was originally derived in Ref.~\cite{Ando1}, 
from which we can write the second-quantized form in the momentum space as: 
\begin{eqnarray}
V_{\mathrm{ep}} = g\sum_{k,q}c_{k+q}^{\dag}\bm{M} c_k (d_q+d_{-q}^{\dag}),
\label{eq1}
\end{eqnarray}
where $c_k = [a_k\;\;b_k]^{\mathrm{T}}$ is the two-component electron annihilation operator in the momentum space for the A and B sublattices and $d_q$ is the phonon annihilation operator. Phonon modes in graphene couple neighbouring A-sublattice and B-sublattice carbon atoms through bond stretching and bending, so that the e-ph coupling becomes an off-diagonal matrix in the pseudo-spin space \cite{Ando1} $g\bm{M}$, where the constant $g = -(\beta v/b^2)\sqrt{\hbar/2NM_c\omega_0}$ gives the magnitude of the e-ph coupling, with $\omega_0 = 0.196\mathrm{eV}$ the optical phonon frequency for graphene from Raman scattering experiments \cite{SSC,Raman}, $N$ the number of unit cells, $M_c$ the mass of a carbon atom, $b = a/\sqrt{3}$ the equilibrium bond length between adjacent carbon atoms and $\beta = \mathrm{d}\,\mathrm{ln}\gamma_0/\mathrm{d}\,\mathrm{ln}b \sim 2$ is a dimensionless parameter that gives the change of the nearest-neighbour tight-binding matrix element $\gamma_0$ with respect to the bond length $b$ \cite{Ando1}. The matrix $M$, for LO/TO phonons, is 
\begin{eqnarray}
\bm{M}(\bm{q}) = \left[\begin{array}{cc}
0 & M_{{\mathrm{AB}}}e^{-i\phi_q} \\
M_{{\mathrm{BA}}}e^{i\phi_q} & 0
\end{array}\right],
\label{eq2}
\end{eqnarray}
with $M_{\mathrm{AB}} = -1$ or $i$ and $M_{\mathrm{BA}} = 1$ or $i$ for LO or TO phonons, respectively, and $\phi_q = \mathrm{tan}^{-1}(q_y/q_x)$ the azithmuthal angle of the momentum $\bm{q}$. We use this model e-ph interaction in our analysis.   

In the leading order theory the e-ph coupling $V_{\mathrm{ep}}$ could be transformed through a canonical transformation to an equivalent phonon-mediated e-e interaction $V_{\mathrm{ee}}^{\mathrm{ph}}$. It follows from Eqs.~(\ref{eq1})-(\ref{eq2}) that the e-e scattering via phonon emission/absorption will flip the sublattice label from A to B or vice versa, and we obtain the following expression for the phonon-mediated e-e interaction resulting from the e-ph interaction of Eq.~(\ref{eq1}) (we adopt the convention where summation over repeated indices is implied): 
\begin{eqnarray}
V_{\mathrm{ee}}^{\mathrm{ph}} = \frac{1}{2}g^2\sum_{k_1,k_2,q}\mathcal{D}^{0}(q,\tau_2-\tau_1)M_{\alpha'\alpha}(\bm{q})M_{\beta'\beta}(-\bm{q}) 
\nonumber \\
c_{k_1+q\alpha'}^{\dag}c_{k_2-q\beta'}^{\dag}c_{k_1\alpha}c_{k_2\beta},
\label{eq3}
\end{eqnarray}
where $\mathcal{D}^{0}(q,\tau_1-\tau_2)$ is the non-interacting retarded phonon Green function in the time domain. In the frequency domain, it is given in terms of the Matsubara bosonic frequency $iq_n$ as $\mathcal{D}^{0}(q,iq_n) = {2\omega_0}/{[(iq_n)^2-\omega_0^2]}$.
%
%
Eq.~(\ref{eq3}) is the central quantity of this paper from which other quantities such as self-energy are derived. To find the self-energy for electrons in the conduction band/holes in the valence band, we work in the chiral basis (i.e. the diagonal basis of the Hamiltonian $H_0$) and denote quantities in the chiral basis with an overhead tilde. We shall focus ourselves only on LO phonons, as the calculations for TO phonons parallel that for LO phonons. The matrix elements for electron scattering from chirality $\lambda \to \lambda'$ and $\mu \to \mu'$ through phonon emission/absorption are  
\begin{eqnarray}
\langle \bm{k}+\bm{q}\lambda'|M(\bm{q})|\bm{k}\lambda\rangle 
&=& \frac{1}{2}[\lambda' e^{-i(\phi_{k+q}-\phi_{q})}-\lambda e^{i(\phi_k-\phi_q)}], \nonumber \\
\langle \bm{k}-\bm{q}\mu'|M(-\bm{q})|\bm{k}\mu\rangle 
&=& \frac{1}{2}[-\mu' e^{-i(\phi_{k-q}-\phi_{q})}+\mu e^{i(\phi_k-\phi_q)}], \nonumber \\
\label{eq5}
\end{eqnarray}
and the corresponding phonon-mediated e-e interaction in the chiral basis can be written as 
\begin{eqnarray}
&&\tilde{V}_{\mathrm{ee}}^{\mathrm{ph}} = \frac{1}{2}g^2\sum_{k_1,k_2,q}\mathcal{D}^{0}(q,\tau_1-\tau_2)\langle \bm{k}_1+\bm{q}\lambda'|M(\bm{q})|\bm{k}_1\lambda\rangle  \nonumber \\
&&\langle \bm{k}_2-\bm{q}\mu'|M(-\bm{q})|\bm{k}_2\mu\rangle 
c_{k_1+q\lambda'}^{\dag}c_{k_2-q\mu'}^{\dag}c_{k_1\lambda}c_{k_2\mu}, 
\label{eq6}
\end{eqnarray}
where the operators $c^{\dag}$, $c$ in Eq.~(\ref{eq6}) denotes the creation and annihilation operators in the chiral basis, with $\lambda, \mu = \pm 1$ the chirality. 
 
The effective many-body velocity/mass renormalization to the band structure comes in large part from the electron-phonon interaction with the Coulomb interaction yielding a quantitatively small correction. 
The effective velocity renormalization due to screened Coulomb interaction was considered in Ref.~\cite{gpFL}. In this Letter, we shall focus on the many-body effects of the e-ph interaction without the effects of Coulomb interaction. 
From Eqs.~(\ref{eq5})-(\ref{eq6}), we have derived the following expression for the electron self-energy in the chiral basis due to the phonon-mediated e-e interaction Eq.~(\ref{eq6}): 
\begin{eqnarray}
\tilde{\Sigma}_{k\pm} &=& -k_BT\frac{g^2}{2}\sum_{\lambda}\sum_{q,iq_n}\mathcal{D}^{0}(iq_n)
\tilde{G}^{0}_{k+q\lambda}(ik_n+iq_n)  \nonumber \\
&&\left[1\mp\lambda\mathrm{cos}(\phi_{k+q}
-2\phi_q)\right], \label{eq7}
\end{eqnarray}
where $\tilde{G}_{k\lambda}^{0}(ik_n) = 1/(ik_n-\xi_{k\lambda})$ is the non-interacting electron Green function in the chiral basis, with $\xi_{k\lambda} = \lambda \epsilon_k-\varepsilon_F$ the quasiparticle energy rendered from the Fermi level. The off-diagonal elements of the self-energy matrix $\tilde{\Sigma}_{k\pm\mp}$, which couples the conduction band and valence band, are found to be zero after performing the angular integration (this is true also for the pure Coulomb interaction case \cite{gpFL}). For TO phonons, Eq.~(\ref{eq7}) remains the same except the angular factor becomes $1\mp\mathrm{cos}(\phi_{k+q}-2\phi_q) \to 1\pm\mathrm{cos}(\phi_{k+q}-2\phi_q)$. The Matsubara sum in Eq.~(\ref{eq7}) can be performed explicitly at zero temperature. For concreteness, we take the case of n-doped graphene and evaluate the self-energy $\tilde{\Sigma}_{k+}$ for quasiparticles in the conduction band (for p-doped material, one should consider $\tilde{\Sigma}_{k-}$ for quasi-holes in the valence band). The quasiparticle decay rate is proportional to the imaginary part of the self-energy, 
%
%
%
\begin{figure}[h]
  \includegraphics[width=8cm,angle=0]{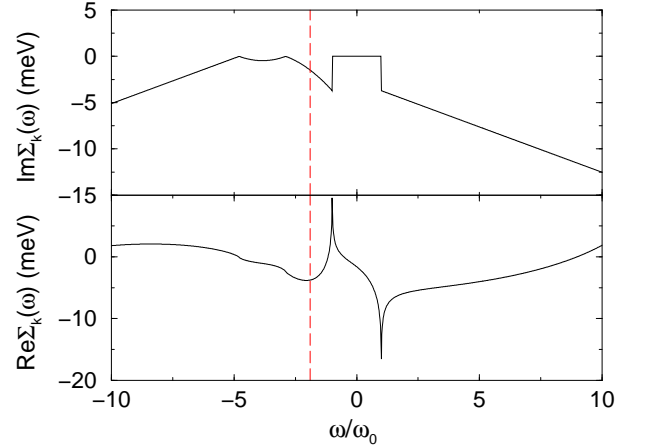} 
\caption{(Color Online) The real and imaginary parts of the self-energy at $k = k_F$ for density $n = 10^{13}\mathrm{cm}^{-2}$. The red dashed line shows the position of the Dirac point. 
} \label{fig1}
\end{figure}
which can be evaluated as: 
\begin{eqnarray}
&&\mathrm{Im}\tilde{\Sigma}_{k+}(\omega) = -\frac{1}{8}g_{\mathrm{ee}}^2 \label{eq16} \\ 
&&\left[F_{\mathrm{intra}}(k,\omega,\omega_0)\theta(\omega-\omega_0)\theta(\omega-\omega_0+\varepsilon_F)\right. \nonumber \\
&&+F_{\mathrm{intra}}(k,\omega,-\omega_0)\theta(-\omega-\omega_0)\theta(\omega+\omega_0+\varepsilon_F) \nonumber \\
&&\left.+F_{\mathrm{inter}}(k,\omega,-\omega_0)\theta(-\omega-\omega_0)\theta(-\omega-\omega_0-\varepsilon_F)\right], \nonumber 
\end{eqnarray}
where the terms 
%
%
$F_{\mathrm{intra}}(k,\omega,\omega_0) = (\omega-\omega_0+\varepsilon_F+\epsilon_k)\left(\omega-\omega_0+\varepsilon_F+\epsilon_k-|\omega-\omega_0+\varepsilon_F-\epsilon_k|\right)/\epsilon_k$ corresponds to the contribution from quasiparticles making an intraband transition in the vicinity of the Fermi level by emitting a phonon, 
and $F_{\mathrm{inter}}(k,\omega,-\omega_0) = -\left\vert \omega+\omega_0+\varepsilon_F+\epsilon_k \right\vert\left(\left\vert\omega+\omega_0+\varepsilon_F+\epsilon_k\right\vert+\omega+\omega_0+\varepsilon_F-\epsilon_k\right)/\epsilon_k$ to quasiparticles making an interband transition from the conduction to the valence band by emitting a phonon, $g_{\mathrm{ee}}^2 \equiv g^2\mathcal{A}/\hbar^2 v^2 = 2\times 10^{-2}$ ($\mathcal{A}$ being the area of the sample) is the dimensionless phonon-mediated e-e coupling constant. 
%
Direct calculation of the real part of the self-energy from Eq.~(\ref{eq7}) gives a logarithmically divergent result, which has been noted as well for the self-energy calculation with only screened Coulomb interaction in the literature \cite{gpFL}. This is because the linear band structure of graphene in the vicinity of the K point is well described by the Dirac Hamiltonian $H_0$, but only up to a cut-off energy scale given by the inverse lattice spacing. The real part of the self-energy should therefore be regularized by introducing the cut-off energy $\Lambda = \hbar v/a$ in the momentum integral in Eq.~(\ref{eq7}), whereupon evaluating the integral we obtain the following exact analytical expression for $\mathrm{Re}\tilde{\Sigma}_{k+}(\omega)$: 
%
\begin{widetext}
\begin{eqnarray}
&&\mathrm{Re}\tilde{\Sigma}_{k+}(\omega) = \frac{1}{4\pi}g_{\mathrm{ee}}^2
\frac{1}{\epsilon_k}\left\{\left\{\theta(k-k_F)\left[-\omega_0\varepsilon_F+(\omega+\omega_0+\varepsilon_F)(\omega+\omega_0+\varepsilon_F+\epsilon_k)\,\mathrm{ln}\left(\frac{\omega+\omega_0+\varepsilon_F}{\omega+\omega_0}\right)\right]\right.\right.\nonumber \\
&&+\theta(k_F-k)\left[-\omega_0\epsilon_k+(\omega+\omega_0+\varepsilon_F)(\omega+\omega_0+\epsilon_k+\varepsilon_F)\,\mathrm{ln}\left(\frac{\omega+\omega_0+\varepsilon_F}{\omega+\omega_0+\varepsilon_F-\epsilon_k}\right)\right. \nonumber \\
&&\left.\left.+(\omega+\omega_0+\varepsilon_F+\epsilon_k)\epsilon_k\,\mathrm{ln}\left(\frac{\omega+\omega_0+\varepsilon_F-\epsilon_k}{\omega+\omega_0}\right)\right]-(\omega_0\to-\omega_0)\right\} \nonumber \\
&&+\epsilon_k(2\omega_0-\epsilon_k)+(\omega_0-\omega_0+\varepsilon_F)(\omega-\omega_0+\varepsilon_F+\epsilon_k)\,\mathrm{ln}\left(\frac{\omega-\omega_0+\varepsilon_F}{\omega-\omega_0+\varepsilon_F-\epsilon_k}\right) \nonumber \\
&&+(\omega_0+\omega_0+\varepsilon_F)(\omega+\omega_0+\varepsilon_F+\epsilon_k)\,\mathrm{ln}\left(\frac{\omega+\omega_0+\varepsilon_F}{\omega+\omega_0+\varepsilon_F+\epsilon_k}\right) \nonumber \\
&&\left.+(\omega-\omega_0+\varepsilon_F+\epsilon_k)\epsilon_k\,\mathrm{ln}\left(\frac{\omega-\omega_0+\varepsilon_F-\epsilon_k}{\omega-\omega_0+\varepsilon_F-\Lambda}\right)+(\omega+\omega_0+\varepsilon_F+\epsilon_k)\,\epsilon_k\mathrm{ln}\left(\frac{\omega+\omega_0+\varepsilon_F+\epsilon_k}{\omega+\omega_0+\varepsilon_F+\Lambda}\right)\right\}, \label{eq17}
\end{eqnarray}
\end{widetext}
In Fig.~\ref{fig1} we show the real and the imaginary parts of the self-energy. The gap in $\mathrm{Im}\tilde{\Sigma}_k(\omega)$ between the phonon energies $-\omega_0$ and $\omega_0$, characteristic of the optical phonon emission/absorption process, results from the Pauli blocking by electrons located within $\omega_0$ of the Fermi level, 
so that decay by an electron with energy $\omega \in [-\omega_0,\omega_0]$ is forbidden because there is no available final state to decay to. Beyond the gap, $\mathrm{Im}\tilde{\Sigma}_k(\omega)$ behaves linearly as $\omega$ for large $\omega/\omega_0$ on both sides of the gap due to the linear dependence on energy of the graphene density of states. At $\omega < -\omega_0-\varepsilon_F$, phonon emission can occur through interband transitions, serving as an extra decay channel in graphene. The interband contribution $F_{\textrm{inter}}$ has two dips in $\vert\textrm{Im}\tilde{\Sigma}\vert$ below the Dirac point (Fig.~\ref{fig1}), one at $-\omega_0-\varepsilon_F$ which comes from phase-space restrictions marking the onset of the interband transition and the other at $-\omega_0-\epsilon_k-\varepsilon_F$ which originates from the angular-dependent electron-phonon interaction vertex Eq.~(\ref{eq5}). 
In regular semiconductors/metals with a parabolic energy dispersion, the imaginary part of the self-energy is often calculated under the approximation of a constant density of states resulting in a square well-shaped profile \cite{Engelsberg},
which suffices to model the quasiparticle decay rate quite accurately for these materials. For graphene, however, the linear $\omega$-dependence of $\mathrm{Im}\Sigma_k(\omega)$ away from the gap is a peculiar feature and we believe that a correct modeling of the existing experimental data for the graphene quasiparticle decay rate \cite{ARPES} should take this feature into account. In Fig.~\ref{fig2}, we show the calculated spectral function $A_k(\omega) \equiv -2\mathrm{Im}\tilde{G}_{k+}(\omega)$, where $\tilde{G}_{k+}(\omega) = 1/[\omega-\xi_{k+}-\tilde{\Sigma}_{k+}(\omega)]$ is the interacting electron Green function with self-energy correction. The spectral function comprises two contributions: a delta function peak within the gap, and a background beyond the gap. As the value of $k$ increases (decreases), the delta function approaches $\omega_0$ ($-\omega_0$) asymptotically from below (above), whereas the background peak at $\omega > 0$ ($\omega < 0$) becomes increasingly prominent.
%
%
\begin{figure}[t]
  \includegraphics[width=7.5cm,angle=0]{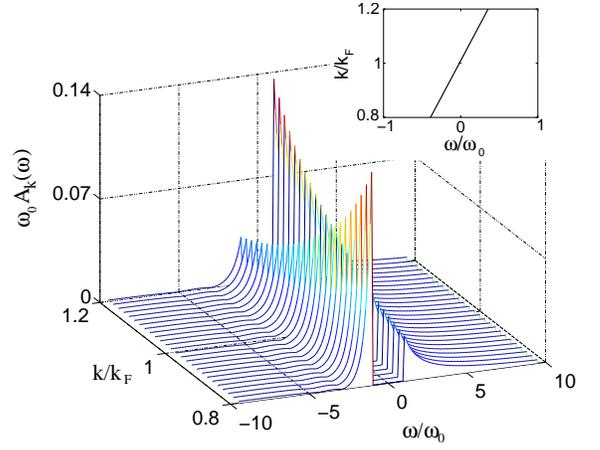} 
\caption{(Color online) The spectral function $A_k(\omega)$ (normalized to dimensionless unit by multiplying $\omega_0$) as a function of $\omega/\omega_0$ and $k/k_F$ for density $n = 10^{13}\mathrm{cm}^{-2}$, the inset shows the locus of the position of the quasiparticle delta function inside the gap $\omega/\omega_0 \in [-1, 1]$.} \label{fig2}
\end{figure}

The derivative of the real part of the self-energy with respect to $\omega$ and $k$ determines the effective velocity $v^*$, which is given by $v^*(k)/v = (1+B)/(1-A)$, where $A = \partial \mathrm{Re}\tilde{\Sigma}_k(\omega)/\partial \omega\vert_{\omega = v(k-k_F)}$ and $B = (1/v)\partial \mathrm{Re}\tilde{\Sigma}_k(\omega)/\partial k\vert_{\omega = v(k-k_F)}$. In metals with a parabolic band dispersion, it is a usual practice to ignore $B$ since in metals $\varepsilon_F/\omega_D \sim 10^{2}$ ($\omega_D$ is the Debye energy) and $B \sim \Sigma/\varepsilon_F \ll A \sim \Sigma/\omega_D$. In graphene, however, $\varepsilon_F/\omega_0 \sim 1$ and $B$ should not be neglected in calculating $v^*/v$. Although the phonon-mediated e-e coupling (whose magnitude is given by $g_{\mathrm{ee}}^2 \sim 10^{-2}$) is in general weaker than the Coulomb coupling (given by the interaction parameter \cite{gpFL} $r_s \sim 0.7$), e-ph interaction actually contributes more significantly to the effective velocity renormalization than Coulomb interaction, because the real part of the self-energy due to e-ph interaction exhibit sharp changes near the phonon energies which are not present in the case of Coulomb interaction. This results in a larger value of the energy derivative of $\mathrm{Re}\tilde{\Sigma}$ near the phonon energy and therefore larger value of $v^*/v$. 
\begin{figure}[t]
  \includegraphics[width=10cm,angle=0]{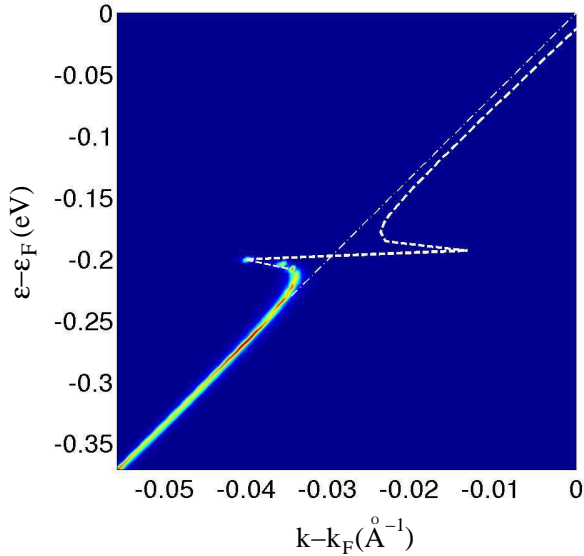}
\caption{(Color online) The renormalized conduction band energy spectrum for $n = 10^{13}\mathrm{cm}^{-2}$. The colored intensity plot shows the magnitude of the spectral function $A_k(\omega)$ whereever it is non-zero, while the dashed white line shows the region within the amount of phonon energy $\omega_0 = 0.196\mathrm{eV}$ from the Fermi level where $A_k(\omega) = 0$. The thin dot-dashed straight line shows the bare unrenormalized spectrum.} \label{fig3}
\end{figure}
\begin{figure}[t]
  \includegraphics[width=5.5cm,angle=270]{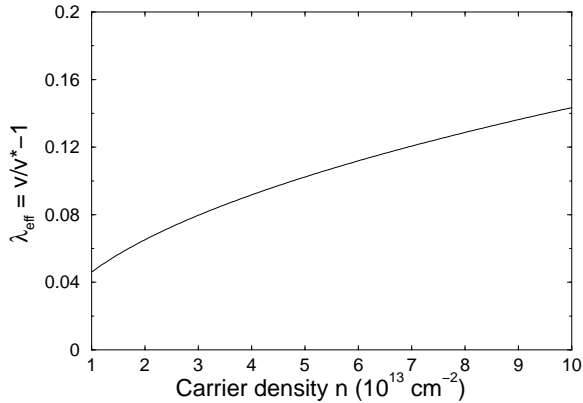} 
\caption{The effective electron-phonon ``coupling'' parameter $\lambda_{\mathrm{eff}} = v/v^*(k_F)-1$ as a function of doping $n$.} \label{fig4}
\end{figure}
In Eq.~(\ref{eq17}), logarithmic singularities occur in $\mathrm{Re}\tilde{\Sigma}_k(\omega)$ at $\pm\omega_0$ where $\mathrm{Im}\tilde{\Sigma}_k(\omega)$ goes through a finite step jump, yielding also logarithmic singularities in the derivatives of $\mathrm{Re}\tilde{\Sigma}_k(\omega)$ with respect to $\omega$ and $k$. Fig.~\ref{fig3} shows the calculated renormalized energy spectrum for electron densities $n = 10^{13} \mathrm{cm}^{-2}$, the sharp kink shows the logarithmic singularity at $\omega = \omega_0$. Such a kink only occurs in the conduction band if the phonon energy is within the Fermi sea $\omega_0 < \varepsilon_F$; if the phonon energy lies outside of the Fermi sea $\omega_0 > \varepsilon_F$ the conduction band will be smooth and the logarithmic singularity could occur in the valence band.  
Although our theory predicts the correct position of the kink (Fig.~\ref{fig3}) in agreement with the experiment \cite{ARPES}, the observed energy spectrum differs from ours in that the sharpness of the kink in the experiment is greatly reduced. First, we note that our zero-temperature theory works well in the temperature regime of the experiment $T < 30\mathrm{K}$ which is much smaller than the Fermi temperature $T_F \sim 4300\mathrm{K}$ at $n \sim 10^{13}\mathrm{cm}^{-2}$, hence smearing of the kink due to finite temperature effect should be minor. We believe that the kink in the experimental spectrum is suppressed due to the combined effect of disorder and screening -- in particular, disorder effects are considerable in the currently existing graphene samples. The combination of these effects will remove the logarithmic singularity leading to a much smoother kink in the energy spectrum, bringing the calculated spectrum in closer agreement with the observed spectrum. Fig.~\ref{fig4} shows the effective electron-phonon ``coupling'' parameter 
$\lambda_{\mathrm{eff}} \equiv v/v^*(k_F)-1$ as a function of electron density $n$. Our results agree in order of magnitude with the extracted value of $\lambda_{\mathrm{eff}}$ from the experiment (Fig.~3c in the second reference of Ref.~\cite{ARPES}), and the band velocity is shown to be reduced by a percentage of $(v-v^*)/v \sim 4-13\%$ from $n = 10^{13}-10^{14}\mathrm{cm}^{-2}$. 

In reality, direct e-e Coulomb interaction, $V_{\mathrm{ee}}^{\mathrm{c}}$, also contributes to the velocity and band renormalization \cite{gpFL} without, however, contributing any kink structure to the energy dispersion. Since both $V_{\mathrm{ee}}^{\mathrm{ph}}$ and $V_{\mathrm{ee}}^{\mathrm{c}}$ are weak, the leading-order theory including both interactions \cite{Jala} would be additive, leading to a ${\lambda}_{\mathrm{eff}} = \lambda_{\mathrm{eff}}^{\mathrm{ph}}+\lambda_{\mathrm{eff}}^{\mathrm{c}}$, where $\lambda_{\mathrm{eff}}^{\mathrm{ph}}$ is shown in Fig.~4 and $\lambda_{\mathrm{eff}}^{\mathrm{c}}$ is considered in Ref.~\cite{gpFL}. We therefore conclude that the graphene velocity renormalization, taking into account both e-ph and Coulomb interaction effects, is of the order of $10-20\%$ aside from the phonon-induced kink at the phonon energy (i.e. Fig.~3).


In summary, we have developed a theory for the phonon-mediated e-e interaction in graphene. We calculated the electron self-energy, the spectral function and the band velocity renormalization due to the e-ph interaction. Our results show that the e-ph coupling has a large effect on the band structure renormalization, exhibiting a kink at $\sim 200\mathrm{meV}$ below the Fermi surface as observed in the experiment and a reduction of the band velocity by $\sim 10\%-20\%$ at the experimental doping level. 


We are grateful to Ben Y.-K. Hu for useful discussion. This work is supported by US-ONR.

Note added: During the preparation of our manuscript, two preprints arXiv:0707.1467v1 by Calandra \textit{et al.} and arXiv:0707.1666v2 by Park \textit{et al.}, appeared dealing with e-ph interaction effects in graphene.


\begin{thebibliography}{18}
\bibitem{SSC} See, for example, the Graphene special issue of Solid State Communications, vol.\textbf{143}, p.1-123 (2007), edited by S. Das Sarma, A.K. Geim, P. Kim, and A.H. MacDonald.
\bibitem{ARPES} A. Bostwick, \textit{et al.}, Nature Phys. \textbf{3}, 36 (2007); J.L. McChesney, \textit{et al.}, arXiv:0705.3264v1. 
\bibitem{Ando1} H. Suzuura and T. Ando, Phys. Rev. B \textbf{65}, 235412 (2002); T. Ando, J. Phys. Soc. Jpn. \textbf{75}, 124701 (2006).
\bibitem{Raman} A.C. Ferrari \textit{et al}., Phys. Rev. Lett. \textbf{97}, 187401 (2006). 
\bibitem{gpFL} S. Das Sarma, E.H. Hwang, and W.-K. Tse, Phys. Rev. B \textbf{75}, 121406(R) (2007); E.H. Hwang, Ben Yu-Kuang Hu, and S. Das Sarma, arXiv:cond-mat/0612345 and arXiv:cond-mat/0703499; Y. Barlas \textit{et al.}, Phys. Rev. Lett \textbf{98}, 236601 (2007).
\bibitem{Engelsberg} S. Engelsberg and J.R. Schrieffer, Phys. Rev. \textbf{131}, 993 (1963).
\bibitem{Jala} R. Jalabert and S. Das Sarma, Phys. Rev. B \textbf{40}, 9723 (1989).

\end{thebibliography}
\end{document}